\documentclass[aps,pra,showpacs,superscriptaddress]{revtex4}
\usepackage[dvips]{graphics}
\usepackage{amsmath,amsfonts,amssymb,amsthm}

\newcommand{\ket}[1]{\mathop{\left|#1\right>}\nolimits}       
\newcommand{\Tr}[1]{\mathop{{\mathrm{Tr}}_{#1}}}              
\newcommand{\Ad}[1]{\mathop{{\mathrm{Ad}}(#1)}}              

\newcommand{\ketbra}[2]{| #1\rangle\!\langle #2 |}
\newcommand{\nn}{\nonumber}
\newcommand{\bbC}{{\mathbb C}}
\newcommand{\bbZ}{{\mathbb Z}}
\DeclareMathOperator*{\arccot}{arccot}

\begin{document}

\title{Relativistically covariant state-dependent cloning of photons}

\author{K. Br\'adler}
\affiliation{Instituto de F\'{\i}sica, Apdo. Postal 20-364,
M\'exico 01000, M\'exico}
\affiliation{School of Computer Science, McGill University,
Montreal, Quebec, Canada}
\author{R. J\'auregui}
\affiliation{Instituto de F\'{\i}sica, Apdo. Postal 20-364,
M\'exico 01000, M\'exico}

\email{kbradler@epot.cz, rocio@fisica.unam.mx}

\date{\today}

\begin{abstract}
The influence of the  relativistic covariance requirement  on the optimality of the symmetric state-dependent $1\to2$ cloning machine is studied. Namely, given a photonic qubit whose basis is formed from the momentum-helicity eigenstates, the change to the optimal cloning fidelity is calculated taking into account the Lorentz covariance unitarily represented by Wigner's little group. To pinpoint some of the interesting results, we found states for which the optimal fidelity of the cloning process drops to $2/3$ which corresponds to the fidelity of the optimal classical cloner. Also, an implication for the security of the BB84 protocol is  analyzed.
\end{abstract}


\maketitle

\section{Introduction}

In recent years the influence of special and general relativity on quantum information processing has begun to be investigated \cite{review_rel}. Staying just within the realm of special
relativity, one of the natural questions is how  Lorentz
transformations affect the properties of both massive and
massless particle states. The basic approach is through
Wigner's little group machinery~\cite{wigner} as the Poincar\'e
group is non-compact and thus without any finite-dimensional unitary
representations~\cite{tung,weinberg}. In this context,
 the entanglement properties of bipartite
states~\cite{first} composed of massive as well as massless
particles have been extensively analyzed.
The fundamental impact on entanglement when the owners of both entangled
subsystems are Lorentz transformed~\cite{sharp_ent_states} has been recognized
since, for physically plausible states, entanglement depends on the properties
of the frame where it is measured~\cite{spread_ent_states,spread_ent_states1}.

In this paper, we go back to single-qubit transformations and
discuss quantum cloning from the relativistic point of view. We
study how the requirement of Lorentz covariance affects the
optimality of a cloning protocol. Quantum cloning has come a long
way since the discovery of the no-cloning theorem~\cite{nocloning}
and one can find an extensive variety of cloners in two recent
review articles~\cite{review_clone}. Lorentz covariance means that
the particular cloning map must be equally effective irrespective of
how any input state is rotated or boosted. More precisely, choosing
the fidelity between an input and output state as a figure of merit
to measure the quality of the clones, we demand that its value be
maximal and independent on the input qubit. The additional
requirement of maximality provides an optimal cloner. As a striking
example of how the relativistic covariance constraint modifies the
optimality of the fidelity results, we investigate the state
dependent $1\to2$ cloner of, generally non-orthogonal, qubits
presented in Ref.~\cite{clone_statedep}. In the relativistic domain
it is necessary to distinguish between particular particle states
for which the effect of the little group generally differ. Photons polarization defines the logical basis of qubits and it is a natural choice due to their use in quantum communication
protocols such as BB84. There are
relatively few previous studies devoted to relativistic effects in either
classical or quantum channels. For the classical ones we highlight
Ref.~\cite{JarettANDCover} where the
channel capacity between two moving observers is studied. For
quantum channels,  there has been a recent growing interest on
quantum information  processing in black holes~\cite{quantumholes}
as well as on how the Unruh effect assists quantum state
encryption~\cite{Unruh}.

\section{Wigner phase and photonic wave packets}

As usual, a standard momentum light-like 4-vector $k^\mu$ with
$k^0>0$ and $k_\mu k^\mu=0$ is chosen. We can transform this vector
into an arbitrary light-like 4-vector $p^\nu=L(p)^\nu_\mu k^\mu$ by
a standard Lorentz transformation. The most general little group
element (stabilizer subgroup which leaves $k^\nu$ invariant) is
$W(\Lambda,p)=L^{-1}_{\Lambda p}\Lambda L_p$~\cite{weinberg} and
consists of rotations and/or translations in a plane, a group which is
isomorphic to the Euclidean group ($ISO(2)$).

The corresponding  Hilbert space is spanned by vectors
with two indices since, together with the angular momentum $\hat
{\bf J}$, the translation operator of the Poincar\'e group $\hat
{\bf P}$ yields the complete set of commuting operators. In any
given reference frame, a rotation around the direction defined by
the standard vector induces a phase on the corresponding state in
Hilbert space:
\begin{equation}
e^{-i\gamma \hat J_3} \vert k;\sigma\rangle =
e^{-i\sigma\gamma}\vert k;\sigma\rangle,
\end{equation}
where for simplicity we took $k_\mu = \omega_0 (1,0,0,1)$ and
$\sigma$ is the component of the angular momentum in the direction
of $k_\mu$ (helicity). As is well known~\cite{weinberg}, massless
particles have only integer or semi-integer $\sigma$ values and for
photons $\sigma=\pm 1 $ holds.

Any other state is obtained from the reference state $\vert
k;\sigma\rangle$ by applying a standard Lorentz transformation.
Thus, if $p_\mu$ is obtained from $k_\mu$ by a rotation with
longitudinal angle $\theta$ and azimuthal angle $\varphi$, then
\begin{equation}
\ket{p;\sigma} = e^{-i\varphi \hat J_3}e^{-i\theta \hat
J_2}e^{i\varphi \hat J_3} \ket{k;\sigma}
\end{equation}
Transforming this state vector by an arbitrary Lorentz
transformation $\Lambda$ we arrive at the unitary representation,
which turns out to be  one-dimensional
$D(W(\Lambda,p))=\exp(i\sigma{\vartheta_W}(\Lambda,p))$. Here
${\vartheta_W}(\Lambda,p)$ is an angle of the rotation dependent on
the Lorentz transformation $\Lambda$ and the initial 4-vector $p$
(the explicit form of ${\vartheta_W}$ can be found
in~\cite{mukunda,wignerphase}). Then, we have
\begin{equation}\label{unitary_repre}
    U(\Lambda)\ket{p,\sigma}=\exp(i\sigma{\vartheta_W})\ket{\Lambda
    p,\sigma}.
\end{equation}

We suppose that a wave packet is prepared in a state
\begin{equation}\label{wavepacket}
\ket{\Psi_f}=\int\sum_{\sigma=\pm1}d\mu(p)f_\sigma(p)a^\dagger_{p,\sigma}\ket{vac},
\end{equation}
where $d\mu(p)$ is a Lorentz-invariant measure and $f_\sigma(p)$ is
a normalized weight function $\int\sum_\sigma
d\mu(p)|f_\sigma(p)|^2=1$ that describes the superposition of modes
with different frequencies but a common direction of propagation
$p$. This selection is made to avoid problems coming from
diffraction effects occurring for a general wave packet so that one
cannot simply define a polarization matrix~\footnote{Considering a
general wave packet with a spatial distribution of momenta, a
general Lorentz transformation yields an intrinsic entanglement
between momenta and polarization degrees of
freedom~\cite{review_rel}. The wider angular spread the packet has
the more severe the influence of this entanglement is on the
definition of the polarization matrix since such an object is not
even rotationally invariant~\cite{polarization}. The second
consequence is that presumably orthogonal helicity states cannot be
perfectly distinguished. If in a given frame the angular and
frequency spreads satisfy $\Delta_{ang}\ll\Delta_{\omega}$, then to keep
the validity of the sharp inequality we have to limit the
distribution of Lorentz boosts. This restriction does not affect the
phase distribution discussed here.}.

Let us examine the action of an arbitrary Lorentz transformation
on~$\ket{\Psi_f}$. We know~\cite{wignerphase} that the phase angle
does not depend on the magnitude of $p$ but just on its direction.
So making the transformation~$\Lambda\ket{\Psi_f}$ the phase
$\exp(i\sigma{\vartheta_W}(\Lambda,p))$ is common for the whole wave
packet. Considering the choice of our wave packet and also the
discussion in~\cite{doppler}, after the Lorentz transformation and
tracing over the momenta degree of freedom, we get
\begin{equation}\label{wavepacket_transf}
\ket{\Lambda\Psi_f}=\int
d\mu(p)\sum_{\sigma=\pm1}e^{i\sigma{\vartheta_W}} f_\sigma(\Lambda
p)\ket{\Lambda
p,\sigma}\stackrel{\Tr{}_p}{\to}\varrho=\begin{pmatrix}
  |\alpha|^2 & \alpha\beta^* e^{2i{\vartheta_W}} \\
  \alpha^*\beta e^{-2i{\vartheta_W}} & |\beta|^2 \\
\end{pmatrix},
\end{equation}
where $|\alpha|^2=\int d\mu(p)|f_1(\Lambda p)|^2$, $|\beta|^2=\int
d\mu(p)|f_{-1}(\Lambda p)|^2$ and $\alpha\beta^*=\int
d\mu(p)f_1(\Lambda p)f_{-1}^*(\Lambda p)$. The helicity basis is the
logical basis $\{\ket{0},\ket{1}\}$ for our qubits (we thus do not
use the Lorentz invariant logical basis composed of two physical
photons proposed in~\cite{relinvinfo} - the task is to clone an
unknown single-photon state).

\section{Relativistically covariant cloning}

For the rest of the article, we assume the following spacetime
arrangement. In her reference frame, Alice prepares a state which
travels in the $p$-direction. Although this direction is well-defined by the outgoing state, for a subject in another inertial
reference frame who receives the state (Bob) it is not sufficient
information. The reason is that there is the whole group of transformations
(rotations around the $p$-direction) which leaves the given
light-like vector intact. This is exactly the 'rotational' part of
Wigner's little group responsible for inducing the Wigner phase
${\vartheta_W}$ in~(\ref{unitary_repre}) and both angles (rotation
and Wigner phase) coincide~\cite{wignerphase}. We  consider
Bob's rotation to be completely unknown and uniformly distributed.

Let us proceed to analyze how the state-dependent cloning setup
investigated by Bru\ss~{\it et al.}~\cite{clone_statedep} is
affected if relativistic covariance is incorporated. First, let us
remember the original problem and later we formulate how relativistic
covariance enters the game. From now on $\sigma_X,\sigma_Y$ and $\sigma_Z$ denote the
Pauli $X,Y$ and $Z$ matrices.

\begin{figure}[t]
\resizebox{10cm}{7cm}{\includegraphics{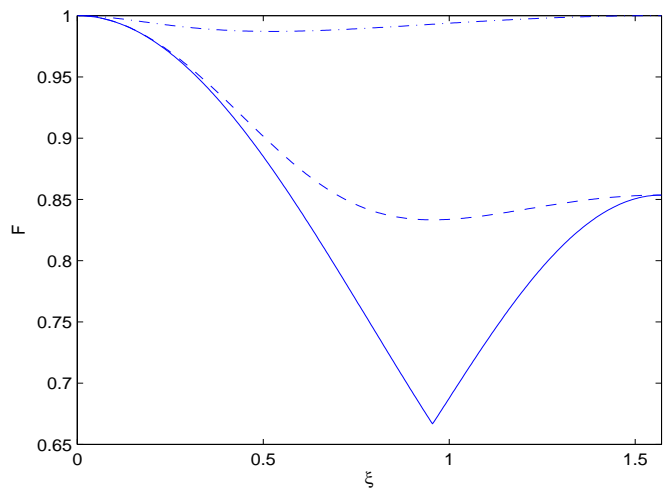}}
\caption{\label{fig_clone_pc} Illustration of how the local
fidelity of a state dependent $1\to2$ symmetric qubit cloner
studied in~\cite{clone_statedep} (dash-dotted line) changes when
some additional symmetries stemming from the relativistic
covariance are required. First, a state dependent phase-covariance
is added (dashed line) and the minimal fidelity
$F^{(1)}_{min}=5/6$ is reached for $\xi_{min}=\arccot\sqrt{1/2}$.
Furthermore, the transposition transformation
corresponding to the finding of an orthogonal complement is
considered and for the same $\xi_{min}$ the minimal fidelity (solid line)
reaches $F^{(2)}_{min}=2/3$.}
\end{figure}

The original problem solved in~\cite{clone_statedep} is, in some
sense, an opposite extreme compared to the universal cloner~\cite{clone},
where all possible pure states are distributed according to the Haar
measure. Here, for a fixed $\xi$, one of just two real states
$\{\ket{\psi}=\cos(\xi/2)\ket{0}+\sin(\xi/2)\ket{1},\sigma_X\ket{\psi}\},\xi\in(0,\pi/2)$
is prepared at random. The authors solved the
problem assuming several reasonable invariance constraints. The
output states were symmetric with respect to the bit flip $\sigma_X$ and
were also permutationally invariant. For later comparison with our
results, the fidelity function obtained in
Ref.~\cite{clone_statedep} is shown in Fig.~\ref{fig_clone_pc}. The
angles of the qubit parametrization are re-scaled to conform the
parametrization used here.

Now we evaluate (and later generalize) the same setup when the relativistic covariance
condition is imposed. Eq.~(\ref{wavepacket_transf}) describes the
effect of Lorentz transformation on the photonic states that are
considered here. On the Bloch sphere, this transformation becomes
$P_{\vartheta_W}=\exp\left(i{\vartheta_W}/2(\openone-\sigma_Z)\right)$.
Since our requirement is that the actual angle of the rotation is
unknown and uniformly distributed so are the states on the Bloch
sphere. Hence, in addition to the symmetries described in the
previous paragraph, we require the invariance of the output
with respect to the operator $P_{\vartheta_W}$. The invariance reflects
the ignorance of the rotation angle that induces the phase angle
${\vartheta_W}$ in~Eq.~(\ref{wavepacket_transf}).

Let us pause here and describe the physical situation. We
suppose that Alice's covariant operation is $\sigma_X$ (plus some
additional operations which we won't mention again) which is
combined with another covariant operation $P_{\vartheta_W}$ induced
by the Wigner phase $\vartheta_W$ (that is, Alice sends one of two
possible states which could be transformed by Bob's rotation) so we
need to compare the action of $P_{\vartheta_W}$ and
$P_{\vartheta_W}\sigma_X$.  The order is important because the
operators do not commute. It can be easily shown that for single
qubits,
\begin{equation}\label{covN1}
    P_{\vartheta_W}\sigma_X=\sigma_XP_{-\vartheta_W}
\end{equation}
what will prove to be very useful for later calculations.

If we want to go beyond the setup studied in~\cite{clone_statedep}
and suppose that Alice may prepare a general pure qubit in the form
$\ket{\psi}_{gen}=\cos(\xi/2)\ket{0}+e^{i\phi}\sin(\xi/2)\ket{1},
\xi\in(0,\pi/2),\phi\in(0,2\pi)$ we find that  $P_{\vartheta_W}$ and
$P_{\vartheta_W}\sigma_X$ (our covariant operations) have a curious
behavior since when they are applied to $\ket{\psi}_{gen}$ these
transformations appear in general as two  asymmetric oriented arcs
 on opposite hemispheres (parametrized by
${\vartheta_W}$). To get a symmetric relativistic transformation we
have to assume a different covariant operation, namely
$\Ad{P_{\vartheta_W}}\varGamma\Ad{\sigma_X}$ where $\Ad{U}[\varrho]=U\varrho U^{-1}$ is the conjugation operation so we are in an adjoint representation of a group whose members are $U$~\cite{Liegroup}.
$\varGamma$ is the transposition of the density matrix in the
standard (logical) basis $\varrho\stackrel{\varGamma}{\to}\varrho^T$ (because of this transformation we traveled into the adjoint representation). The reason for incorporating $\varGamma$
becomes evident when we compare the action of $\Ad{P_{\vartheta_W}}$ and
$\Ad{P_{\vartheta_W}}\varGamma\Ad{\sigma_X}$ (our new covariant couple)
on $\ket{\psi}_{gen}$. In this case, we will make use of the
following identity (see proof in Appendix)
\begin{equation}\label{covN2}
     \Ad{P_{\vartheta_W}}\varGamma\Ad{\sigma_X}=\varGamma\Ad{\sigma_XP_{\vartheta_W}}.
\end{equation}
Note that $[\varGamma,\Ad{\sigma_X}]=0$.
The motivation for introducing the identity is purely computational
(just as for commutator~(\ref{covN1})) but the physical interpretation
is interesting.  Since $\left[\Ad{P_{\vartheta_W}},\varGamma\Ad{\sigma_X}\right]=0$
holds the order of the covariance operations does not matter. Thus, in the next we will investigate both relativistic covariance effects, i.e. when covariance is required with respect to
$\Ad{P_{\vartheta_W}\sigma_X}$ for real states $\{\ket{\psi}\}$, and
$\Ad{P_{\vartheta_W}}\varGamma\Ad{\sigma_X}$ for general states
$\{\ket{\psi}_{gen}\}$. Except where really necessary, we will omit the symbol $\Ad{}$ for the conjugation operation to avoid the excessive notation but we have to remember that we keep working in the adjoint representation.

Let us rephrase the invariance requirements
from the previous paragraph in an appropriate formalism. The Jamio\l
kowski isomorphism~\cite{jamiolk} between positive operators and CP
maps~\cite{dariano+presti} is a traditional tool for the
calculation of optimal and group covariant completely positive
(CP) maps. One appreciates the representation even more by realizing
that an implementation of the mentioned transposition operation is
particularly easy. Let $\mathcal{M}$ be a CP map, then the
corresponding positive operator $R_\mathcal{M}$ is related by
\begin{equation}\label{jamiolk}
     \mathcal{M}^{(N)}(\varrho_{in})
     =\Tr{in}\left[\left(\openone\otimes\varGamma^{\circ
     N}\left[\varrho_{in}\right]\right)R^{(N)}_\mathcal{M}\right],
\end{equation}
with $N=1,2$ denoting the above discussed alternatives (without
and with the transposition, respectively) and
$\varGamma^{\circ1}\equiv\varGamma,\varGamma^{\circ2}=\varGamma\circ\varGamma=\openone$.
The expression  $\varGamma^{\circ 1}\left[\varrho_{in}\right]\equiv\varrho^T_{in}$ stands for the transposition of the density matrix $\varrho_{in}$. It is important
to stress that the case $N=2$ must not be in a contradiction with
the definition of the isomorphism ($N=1$). Consequently, the net
effect is that we require $R^{(2)}_\mathcal{M}$ to be invariant
with respect to the transposition of $\varrho_{in}$.

If $\mathcal{M}$ is a cloning CP~map then using Eq.~(\ref{covN1}) for $N=1$ and Eq.~(\ref{covN2}) for $N=2$ we can first start by requiring covariance with respect to $P_{\mp\vartheta_W}$. Then the covariance conditions in both representations (standard and Jamio\l kowski, respectively) read
\begin{subequations}\label{covcond}
\begin{gather}
\mathcal{M}^{(1)}(\varrho)=(P_{-\vartheta_W}\otimes
P_{-\vartheta_W})^\dagger\mathcal{M}\left(P_{-\vartheta_W}\varrho P^\dagger_{-\vartheta_W}
\right)(P_{-\vartheta_W}\otimes
P_{-\vartheta_W}) \rightleftharpoons
[R^{(1)}_\mathcal{M},P_{-\vartheta_W}\otimes
P_{-\vartheta_W}\otimes P^*_{-\vartheta_W}]=0\label{covcondN1}\\
\mathcal{M}^{(2)}(\varrho)=(P_{\vartheta_W}\otimes
P_{\vartheta_W})^\dagger\mathcal{M}\left(P^*_{\vartheta_W}\varrho^T
P_{\vartheta_W}^T\right)(P_{\vartheta_W}\otimes
P_{\vartheta_W}) \rightleftharpoons
[R^{(2)}_\mathcal{M},P_{\vartheta_W}\otimes
P_{\vartheta_W}\otimes P_{\vartheta_W}]=0.\label{covcondN2}
\end{gather}
\end{subequations}
Note that $P^*_{\vartheta_W}=P^\dagger_{\vartheta_W}=P_{-\vartheta_W}$. Let us explain the use of the Jamio\l kowski isomorphism. For $N=1$, utilizing Eq.~(\ref{covN1}) we apply the covariance condition coming from the structure of the phase operator $P_{-\vartheta_W}$. We get the basic structure of the Jamio\l kowski operator $R^{(1)}_\mathcal{M}$ and we apply the bit-flip and the output state symmetry covariance conditions. Similarly for $N=2$, the strategy is to summon the rhs of Eq.~(\ref{covN2}) to find how the covariance condition coming from $P_{\vartheta_W}$ defines the basic structure of $R^{(2)}_\mathcal{M}$. Then, in addition to the previously mentioned covariant conditions, we require the covariance regarding the transposition of an input state. This is the reason why on the lhs of Eq.~(\ref{covcondN2}) there is $\varGamma\circ[P_{\vartheta_W}\varrho P^\dagger_{\vartheta_W}]=P^*_{\vartheta_W}\varrho^T
P_{\vartheta_W}^T$. Finally, we calculate single-copy
fidelities of the cloned state for both cases. The operator
$R^{(1)}_\mathcal{M}$ is just a unitary modification of
$R^{(2)}_\mathcal{M}$ as seen from~Eqs.~(\ref{covcond}) so it is
sufficient to analyze the structure of the case $N=2$ and for $N=1$ to
subsequently modify the operator by $\sigma_Y\otimes\sigma_Y\otimes\openone$ -- it is a simple permutation of basis states.

One of the Schur lemmas gives us the
structure of the positive operator $R^{(2)}_\mathcal{M}$. It is a
sum of the isomorphisms between all equivalent irreducible
representations, which in the case of $P_{\vartheta_W}\in U(1)$ are
all one-dimensional and are distinguished by the character values
$e^{in{\vartheta_W}}$ with $n\in\bbZ$. More specifically, $P_{\vartheta_W}\otimes
P_{\vartheta_W}\otimes P_{\vartheta_W}$ is composed of four irreducible representations. Two of them are one dimensional (spanned by $\{\ket{0}\},\{\ket{7}\}$) and two are three dimensional $\{\ket{1},\ket{2},\ket{4}\},\{\ket{3},\ket{5},\ket{6}\}$ where $\ket{m}$ is a decimal record of the 3-qubit basis.

Taking into account the above
discussed additional symmetries of $R^{(1,2)}_\mathcal{M}$ the number of independent parameters
gets limited and we arrive to the following form of
$R^{(2)}_\mathcal{M}$
\begin{eqnarray}\label{Roperator}
&R^{(2)}_\mathcal{M}=c_{00}(\ketbra{0}{0}+\ketbra{7}{7})+
c_{11}(\ketbra{1}{1}+\ketbra{6}{6})+
c_{22}(\ketbra{2}{2}+\ketbra{5}{5})+
c_{33}(\ketbra{3}{3}+\ketbra{4}{4})\nonumber\\
&+c_{24}(\ketbra{2}{4}+\ketbra{4}{2}+\ketbra{3}{5}+\ketbra{5}{3})\nonumber\\
&+c_{12a}[(\ketbra{1}{2}+\ketbra{1}{4}+\ketbra{3}{6}+\ketbra{5}{6}+h.c.)
+ic_{12b}(\ketbra{1}{2}+\ketbra{1}{4}+\ketbra{3}{6}+\ketbra{5}{6}-h.c.)],
\end{eqnarray}
where $c_{ij}\in\bbC$ (for $i\not=j$) are coefficients of the isomorphisms
$\ketbra{i}{i}\leftrightarrow\ketbra{j}{j}$  and
$c_{12a}=\Re[c_{12}],c_{12b}=\Im[c_{12}]$.  Two additional
conditions come from the trace-preserving constraint
$\Tr{out}\left[R^{(N)}_\mathcal{M}\right]=\openone\Rightarrow
c_{00}+c_{11}+c_{22}+c_{33}=1$ (common for both $N$) and, of course,
from the positivity condition $R^{(N)}_\mathcal{M}\geq0$.

Since we are cloning a pure qubit, our figure of merit to be
maximized is the single copy fidelity between the input states
$\ket{\psi}=\cos(\xi/2)\ket{0}+\sin(\xi/2)\ket{1}$ (for $N=1$) and
$\ket{\psi}_{gen}=\cos(\xi/2)\ket{0}+e^{i\phi}\sin(\xi/2)\ket{1}$ (for $N=2$) and the target states of the same form
\begin{subequations}\label{fidelity}
\begin{gather}
    F^{(1)}=\Tr{}
    \left[\left(\ketbra{\psi}{\psi}\otimes\openone\otimes\varGamma
    \left[\ketbra{\psi}{\psi}\right]\right)
    R^{(1)}_\mathcal{M}\right]\label{fidelityN1}\\
    F^{(2)}=\Tr{}
    \left[\left(\ketbra{\psi}{\psi}_{gen}\otimes\openone\otimes
    \ketbra{\psi}{\psi}_{gen}\right)
    R^{(2)}_\mathcal{M}\right]\label{fidelityN2}.
\end{gather}
\end{subequations}
Observe that in Eq.~(\ref{fidelityN2}) the transposition operator was additionally applied.

\paragraph*{{\bf Case $N=1$} {\rm (covariance w.r.t. $P_{\vartheta_W}$ and
$P_{\vartheta_W}\sigma_X$)}}

If we apply the covariant operations on an arbitrary real
$\ket{\psi}$ then for different values of the Wigner phase
$\vartheta_W$ we generate two symmetric trajectories on the opposite
hemispheres of the Bloch sphere. Reformulating the search for the
fidelity as a semidefinite program using the SeDuMi
solver~\cite{SeDuMi} in the YALMIP environment~\cite{YALMIP} the
number of parameters is reduced and $R^{(1)}_\mathcal{M}$ can be
diagonalized. This leads to the full analytical derivation of the
fidelity function~(\ref{fidelityN1}) as a function of the input
state $\ket{\psi}$
\begin{equation}\label{fidelanalN1}
    F^{(1)}={1\over2}\left[
    1+{1\over2}\cos^2{\xi}\left(1+{\cos^2{\xi}\over\sqrt{2\sin^4{\xi}+\cos^4{\xi}}}\right)
    +{\sin^4{\xi}\over\sqrt{2\sin^4{\xi}+\cos^4{\xi}}}
    \right].
\end{equation}
The function is depicted in~Fig.~\ref{fig_clone_pc}
and we notice several interesting things. Obviously, the fidelity
is lower than the original state dependent fidelity. We observe
that the minimum moved from $\xi_{min}^{\text
Bru\ss}=\pi/6$~\cite{clone_statedep} to the angle
$\xi_{min}=\arccot\sqrt{1/2}$ but more interesting point is that the
fidelity attains $F^{(1)}_{min}=5/6$. This value is 'reserved' for
the $1\to2$ universal symmetric cloner~\cite{clone}, i.e. the cloning map
covariant with respect to the action of $SU(2)$ (or, equivalently,
to the cloning of all mutually unbiased states of the Bloch
sphere~\cite{review_clone}). Such a low value for a kind of
phase-covariant cloner we are investigating may be surprising. For
$\xi=\pi/2$ we recover the result from~\cite{clone_pc} where
$F={1\over2}+\sqrt{1\over8}$. This is expected because the bit
flip (one of our additional conditions) is unnecessary on the
equator (due to the presence of $P_{\vartheta_W}$).

\paragraph*{{\bf Case $N=2$} {\rm (covariance w.r.t. $P_{\vartheta_W}$ and
$P_{\vartheta_W}\varGamma\circ\sigma_X$)}}

Using  methods similar to those in the previous paragraph we arrive with the help of~Eq.~(\ref{fidelityN2}) and $\ket{\psi}_{gen}$ at the following form of the fidelity function
\begin{equation}\label{fidelanalN2}
    F^{(2)}=\max{\Biggl\{{1\over4}(\cos2\xi+3),{1\over2}\left[
    1+{1\over2}\cos^2{\xi}\left(-1+{\cos^2{\xi}\over\sqrt{2\sin^4{\xi}+\cos^4{\xi}}}\right)
    +{\sin^4{\xi}\over\sqrt{2\sin^4{\xi}+\cos^4{\xi}}}
    \right]\Biggr\}},
\end{equation}
which is independent on the input state phase $\phi$.
This result is no less interesting and the function is again
depicted in~Fig.~\ref{fig_clone_pc}. The minimum angle is common
with the previous case but the corresponding fidelity drops to
$F^{(2)}_{min}=2/3$. This low value can be justified
if we realize what kind of operation corresponds to $N=2$. We
combine two impossible operations, quantum cloning and finding
the universal-NOT operation, into what is together known as
the anti-cloning operation~\cite{anticlone}. This combined requirement is
apparently stronger than the universal (i.e. $SU(2)$) covariance
and the reason for the low fidelity values is that the map
$R^{(2)}_\mathcal{M}$ must be of the same form for both $\varrho$
and $\varrho^T$ as a result of~Eq.~(\ref{jamiolk}).

Could this result tell us something about, for instance, the
security of quantum key distribution (QKD)? Looking at the most
studied protocol BB84~\cite{bb84} (of course, implemented by the
polarization encoding which is preferred for a free-space
communication for which the relativistic effects may be very
relevant) we see that four qubits equidistantly distributed on the
meridian are used. For $N=1$, they form the $xz-$plane of the Bloch
sphere (`real meridian') and for $N=2$ it is an arbitrary grand
circle intersecting the north and south pole (`complex meridian').
From the viewpoint of an eavesdropper without the knowledge of the
Wigner phase ${\vartheta_W}$ and decided to clone the quantum states
to get some information, we can now demonstrate that  not all
quadruples are equally good. If the states
$\{\cos(\pi/8)\ket{0}\pm\sin(\pi/8)\ket{1},\cos(\pi/8)\ket{1}\pm\sin(\pi/8)\ket{0}\}$
are used for the QKD purposes then by inserting $\xi=\pi/4$
into~Eqs.~(\ref{fidelanalN1}) and (\ref{fidelanalN2}), we get
$F^{(1)}={(5+\sqrt{3})/8}\simeq0.8415$ and $F^{(2)}=3/4$. On the
other hand, using the quadruple
$\{\ket{0},\ket{1},1/\sqrt{2}\ket{0\pm1}\}$ we get the fidelity
$F=5/6$ because by phase-rotating the quadruple states (that is,
applying $P_{\vartheta_W}$) we pass the mutually unbiased states of
the Bloch sphere. We see that $F^{(2)}<F<F^{(1)}$ corresponds
to the fact that for $N=2$ the eavesdropper has less information
about the input state.

Another interesting question is how the relativistic covariance affects the
optimality of the universal cloner. Here the situation is different. In the analysis
above we combined two covariant operations ($\varGamma\circ\sigma_X$ and
$P_{\vartheta_W}$) which are not generally subsets of each other.
On the other hand, as we saw, every Wigner rotation
is a $U(1)$ covariant rotation and since $U(1)\subset SU(2)$ we may conclude
that the optimality of the universal cloner will remain unchanged.
Pictorially, it corresponds to the situation where Alice sends a
completely unknown photon ($SU(2)$ covariance) to Bob who, in addition,
does not know how the whole Bloch sphere rigidly rotates
(his rotation with respect to Alice). However, this is again a kind of $SU(2)$ rotation.

\section{Conclusions}

In conclusion, we investigated the role played by the
requirement of relativistic covariance in the problem of the
optimality of one of the most prominent forbidden
quantum-mechanical process as quantum cloning. Observing that the
effect of Wigner's little group can be translated into the
language of so-called phase-covariant processes we studied how
the effectiveness of the cloning process becomes modified.
Particularly, we considered an observer in a different reference
frame with no knowledge of the parameters of the reference
frame where the state designated for cloning was produced. Here we
focused on the class of state-dependent cloners where the effect
is especially appreciable. First, as a direct application of the relativistic
considerations on the cloning setup studied in~\cite{clone_statedep}
where one of two real states is prepared in one inertial frame and cloned
in another inertial frame whose transformation properties regarding
the first one are completely unknown. Second, we went beyond this
setup and supposed that in the first frame
two general pure qubits related by a common action of the Pauli
$X$ matrix and the density matrix transposition operator might be prepared.
Again, we wanted such a state to be cloned in another inertial frame without knowledge of which state was actually sent and how it was relativistically transformed.
In both cases, we brought analytical expressions for local fidelities of the
output states asking the fidelity to be maximal and optimal. One of the intriguing
results is that in the second case the fidelity drops even below the universal
cloner limit. The reason is that we combined the mentioned cloning invariance
conditions with another forbidden process - finding the orthogonal complement
of an unknown state.
Note that even without the relativistic context we generalized the previous
research on the phase-covariant cloning maps and at the same time we studied
optimal covariant processes considering covariance operations which do not commute.

As an example of the consequences for communication security
issues we have shown that for an eavesdropper determined to get some
information by cloning a BB84 quadruple of states, not all
possibilities are equally good and some provide him with more
information.

\begin{acknowledgments}
The authors are grateful to Patrick Hayden for reading the manuscript.
\end{acknowledgments}

\appendix
\section{}
\label{app_covN2}
In the following we use some of the basic properties of Lie groups~\cite{Liegroup}.

Let $A,B,C$ be invertible linear transformations. We define the conjugation operation $c(B)[\varrho]\equiv\Ad{B}[\varrho]=B\varrho B^{-1}$ (and similarly $c(C)$) satisfying $[A,c(B)]=0,[A,c(C)]\not=0,[c(B),c(C)]\not=0$. Then if $A^2=c(B^2)$ we have $A\,c(CB)-c(BC)A=0$.

{\it Proof.} Noting that
\begin{align}\label{proof}
A\,c(CB)-c(BC)A        & = 0        &&         \nn\\
A^2\,c(CB)A-A\,c(BC)A^2  & = 0        && \text{multiplied by $A$ from left and right} \nn\\
A^2\,c(C)D-D\,c(C)A^2    & = 0        && D=A\,c(B)=c(B)A \nn\\
A^{-2}\,D\,c(C)A^2       & = c(C)D.   &&
\end{align}
Similarly, we get $c(B^{-2})D\,c(CB^2)=c(C)D$. Equalling these two expressions we immediately see that $Dc(C)=A^2\,c(B^{-2})D\,c(CB^2)A^{-2}$ holding if $A^2=c(B^2)$.$\Box$
\\
Now we identify $A=\varGamma,c(B)=\sigma_X$ and $c(C)=P_{\vartheta_W}$ so we have $\varGamma P_{\vartheta_W}\sigma_X[\varrho]=\sigma_XP_{\vartheta_W}\varGamma[\varrho]$ and to get Eq.~(\ref{covN2}) we apply $\varGamma$ on the equation from the left and right using the fact that $\varGamma^{\circ2}[\varrho]=\openone$.

\end{document}